# A simple model for in- and out-of-plane resistivities of hole doped cuprates

S. H. Naqib[*1], M. Afsana Azam[1,4], M. Borhan Uddin[1,2], and J. R. Cole[3]

[1]*Department of Physics, University of Rajshahi, Rajshahi-6205, Bangladesh*

[2]*Department of CSE, International Islamic University Chittagong, Bangladesh*

[3]*Cambridge Flow Solutions Ltd., Histon, Cambridge CB24 9AD, UK*

[4]*Department of Physics, DUET, Gazipur, Dhaka, Bangladesh*

**Abstract**

The highly anisotropic and qualitatively different nature of the in- and out-of-plane charge dynamics in high-$T_c$ cuprates cannot be accommodated within the conventional Boltzmann transport theory. The variation of in-plane and out-of-plane resistivities with temperature and hole content are anomalous and cannot be explained by Fermi-liquid theory. In this study, we have proposed a simple phenomenological model for the dc resistivity of cuprates by incorporating two firmly established generic features of all hole doped cuprate superconductors – (i) the pseudogap in the quasiparticle energy spectrum and (ii) the $T$-linear resistivity at high temperatures. This $T$-linear behavior over an extended temperature range can be attributed to a quantum criticality, affecting the electronic phase diagram of cuprates. Experimental in-plane and out-of-plane resistivities ($\rho_p(T)$ and $\rho_c(T)$, respectively) of double-layer Y(Ca)123 have been analyzed using the proposed model. This phenomenological model describes the temperature and the hole content dependent resistivity over a wide range of temperature and hole content, $p$. The characteristic PG energy scale, $\varepsilon_g(p)$, extracted from the analysis of the resistivity data was found to agree quite well with those found in variety of other experiments. Various other extracted parameters from the analysis of $\rho_p(T)$ and $\rho_c(T)$ data showed systematic trends with changing hole concentration. We have discussed important features of the analysis in detail in this paper.

## 1. Introduction

Superconductivity in hole doped cuprates poses an outstanding problem for the scientific community. After almost three decades of its discovery a number of unsolved puzzles exist and a proper understanding of the normal and superconducting (SC) ground states of these strongly correlated electronic systems remains elusive till date [1 – 3]. In high-$T_c$ cuprates, non-Fermi liquid charge transport in the normal state and other anomalous physical properties have provided with many challenging issues and have stimulated significant interest [4 – 7]. One of the key questions to be answered is that why the out-of-plane resistivity differs so much from the in-plane resistivity in the normal state. The resistivity anisotropy ratio, $\rho_c/\rho_p$, shows a temperature dependent behavior that begs explanation. In the underdoped (UD) region, below certain characteristic temperature, $\rho_c(T)$ is "semiconductor" like ($d\rho_c/dT < 0$), in contrast to the in-plane resistivity, $\rho_p(T)$, which is metallic ($d\rho_p/dT > 0$). This marked difference between $\rho_c(T)$ and $\rho_p(T)$ is not what one might expect within the conventional Fermi-liquid theory [6]. Such different temperature dependence and very large (~ $10^5$ in Bi2212 at certain doping) value of the resistivity anisotropy have stimulated vigorous theoretical and experimental investigations on the in-plane and interlayer charge dynamics of high-$T_c$ cuprates and their possible link with superconductivity itself [6, 8 – 10]. However, a complete, empirically relevant, and unified description of the *ab*-plane and *c*-axis charge transport in hole doped cuprates is still lacking.

Experimental studies reveal that the generic features of the temperature and hole content dependences of resistivity are qualitatively identical in all the families of hole doped cuprates, even though structural and electronic anisotropies vary over a large range. Therefore, it is reasonable to assume that the dominant electronic correlations present in all hole doped curates, irrespective of the structural or other finer electronic details (e.g., strength of spin-charge *stripe* correlations [11, 12]) are also responsible for the non-Fermi-liquid behavior temperature dependent resistivity and its evolution with number of added holes, $p$, in the copper oxide planes. Besides superconductivity at high $T_c$ itself, the pseudogap (PG) correlation in the normal state is the single most robust characteristic of all hole doped cuprates. Various physical properties including the resistive features of cuprates are dominated by the PG, especially in the UD regime [13, 14]. Any formalism aimed to explain the $T$ and $p$ dependent resistivity must take into consideration of the full impact of the $p$-dependency of the PG in quasiparticle (QP) energy density of states (EDOS).

A number of theoretical study have been carried out to model the in- and out-of-plane dc charge transport of hole doped cuprates [15 - 20]. Theoretical approaches vary from conventional Fermi-liquid, nonconventional, phenomenological, to outright exotic. It is fair to say that these models have limited success. One major drawback of most of these models is that the theoretical framework used to describe the doping and temperature dependent resistivity does not hold when wider bodies of other experimental facts are considered.

In a previous study [21], we have modeled the *c*-axis resistivity of some bi-layer cuprates by taking into account of the effect of PG on the interlayer charge transport. In this study we wish to extend the model for in-plane charge transport. We show that the essential difference between in- and out-of – plane charge transport lies within the way in which the PG affects in- and out-of-plane charge dynamics. Our proposed model also yields mutually consistent set of values for the characteristic energy scales for the PG from the analysis of in- and out-of-plane resistivity data,

The paper is organized as follows. In section 2, we describe the outline of the proposed model. Next, in section 3, experimental $\rho_p(T)$ and $\rho_c(T)$ data of double-layer Y(Ca)123 single crystals have been analyzed. Discussion on the important findings constitutes section 4 and finally, conclusions are drawn in section 5.

## 2 Phenomenological model

As mentioned in the previous section, *T*-linear resistivity and the ubiquitous effect of PG in the quasiparticle on the charge transport are the two generic features present in all hole doped cuprates. Strange *T*-linear resistivity calls for an unconventional approach and quantum criticality seems to be the most plausible answer, especially when its potential to induce Cooper pairing at high temperature in systems with strong electronic correlations is considered [22].

Ordinary phase transitions are driven by thermal fluctuations and involve a change between an ordered and a disordered state. At absolute zero, where there are no thermal fluctuations, a fundamentally new type of phase transition can occur - a quantum phase transition. Quantum phase transitions are triggered by quantum fluctuations associated with the uncertainty principle. This type of phase transition involves no change in entropy and can be accessed only by varying a non-thermal parameter such as the doped holes as in case of cuprate superconductors. The point that separates the two distinct quantum phases at zero temperature is called a quantum critical point (QCP). The detailed description of how such a QCP affects the temperature dependent charge transport is still lacking. But a *T*-linear resistivity may originate from quantum criticality quite naturally. The basic physics is outlined below.

From critical scaling, it follows that the thermal equilibrium time, $\varGamma_{qcp}$, at a QCP is given by [22, 23]

$$\varGamma_{qcp} = \frac{Ch}{2\pi k_B T} \qquad (1)$$

where $h$ is the Planck's constant, $k_B$ is the Boltzmann's constant and $C$ is a dimensionless universal parameter depending on dimensionality of the system. The most intriguing feature of Eqn. 1 is that this characteristic time depends only on temperature; no other characteristic energy scale (e.g., Fermi energy, exchange energy etc.) relevant to the Fermi-liquid description of the system appears in this expression. The transport properties are drastically affected by the presence of quantum criticality. Because the values of various transport coefficients depend on the same process that establish local thermal equilibrium inside the system under consideration.

At high-$T$ above the QCP, thermal timescale is much shorter than the quantum timescale, the physical properties at finite temperatures are, therefore, seriously influenced by the presence of the QCP at a particular value of the non-thermal parameter (say, $g = g_c$, where $g_c$ is the critical value of the non-thermal parameter). The system in this regime cannot be simply described by the ground state wave function at $g$. In this quantum critical regime, the temperature dependence of the physical quantities often exhibits a striking deviation from conventional Fermi-liquid behavior. As temperature becomes the only relevant energy scale; the scattering rate is given by the inverse of Eqn. 1, and the resistivity becomes completely linear in this region of the $T$-$p$ phase diagram. This $T$-linearity should affect both in- and out-of plane normal state resistivities of cuprates in the same qualitative way.

In general the dc electrical resistivity of a metal is expressed via the Drude formula given by

$$\rho = \frac{m^*}{ne^2\tau} \qquad (2)$$

where $m^*$ is the carrier effective mass, $n$ is the carrier concentration, $e$ being the charge of the carrier and $\tau$ is the relaxation time. For ordinary Fermi-liquids the temperature dependence of $\rho$ arises from the temperature dependent scattering rate, $1/\tau$. PG effects the in-plane resistivity by reducing the carrier scattering rate [24]. The link between the PG in the QP EDOS and the scattering rate can be established using the time-dependent perturbation theory. Using Fermi's golden rule for transition probability, we get

$$\frac{1}{\tau} = \frac{2\pi}{\hbar} N(\varepsilon_F) <i|H'|f>^2 \qquad (3)$$

where $N(\varepsilon_F)$ is the EDOS of the Fermi level and $H'$ is the scattering Hamiltonian, $i$ and $f$ denote the initial and final states of the scattered charge carrier. At finite temperatures the scattering rate will, in fact, be determined by the thermally averaged EDOS at the Fermi-level, $<N(\varepsilon_F)>_T$. To explain a number of diverse physical properties of different families of high-$T_c$ cuprates, $<N(\varepsilon_F)>_T$ has been modeled quite successfully using the following expression [25]

$$<N(\varepsilon_F)>_T = N_0 \left[1 - \left(\frac{2T}{\varepsilon_g}\right)\ln\left\{\cosh\left(\frac{\varepsilon_g}{2T}\right)\right\}\right] \quad (4)$$

where $N_0$ is the flat EDOS outside the PG region and $\varepsilon_g$ is the characteristic PG energy scale expressed in temperature. Thus, considering the effects of QCP and PG, the in-plane resistivity of hole doped cuprates is dominated by two terms, the first one, due to quantum criticality, has the form

$$\rho_{QCP} = \alpha_p T \quad (5)$$

and the second one proportional to the depleted EDOS due to PG is

$$\rho_{PG} = \beta_p N_0 \left[1 - \left(\frac{2T}{\varepsilon_g}\right)\ln\left\{\cosh\left(\frac{\varepsilon_g}{2T}\right)\right\}\right] \quad (6)$$

where $\alpha_p$ is a constant, depending on hole content, measuring the strength of hole scattering and $\beta_p$ is an weight factor taking into account of the effect of the overall momentum dependence of $<N(\varepsilon_F)>_T$ and the scattering matrix elements on in-plane resistivity. Thus the total in-plane resistivity can be expressed as

$$\rho_p(T) = \rho_{0p} + \alpha_p T + \beta_p N_0 \left[1 - \left(\frac{2T}{\varepsilon_g}\right)\ln\left\{\cosh\left(\frac{\varepsilon_g}{2T}\right)\right\}\right] \quad (7)$$

which on rearrangement becomes

$$\rho_p(T) = a_p + \alpha_p T - b_p\left(\frac{2T}{\varepsilon_g}\right)\ln\left\{\cosh\left(\frac{\varepsilon_g}{2T}\right)\right\} \quad (8)$$

with $a_p = (\rho_{0p} + \beta_p N_0)$ and $b_p = \beta_p N_0$. We have used Eqn. 8 to fit the experimental $\rho_p(T)$ data for Y(Ca)123 over a wide range of compositions. It should be noted that we have ignored the possible temperature dependence of the square of the matrix element for transition (Eqn. 3) in the expression for $\rho_p(T)$. Within the proposed scenario the $T$-dependence of the $\rho_{PG}$ comes primarily from the thermally averaged EDOS (Eqn. 4).

As mentioned before, the out-of-plane resistivity has been modeled within the same phenomenology before [21]. For the sake of completeness we give the expression for $\rho_c(T)$ below

$$\rho_c(T) = \alpha_c T + \frac{\beta_c}{[1 - (\frac{2T}{\varepsilon_g})\ln\{\cosh(\frac{\varepsilon_g}{2T})\}]^2} \qquad (9)$$

where, $\beta_c = 1/At_c^2 N_0^2$. Here $t_c$ is the $c$-axis tunneling matrix element and $N_0$ is the flat EDOS at high energies outside the PG region. These two parameters are expected to be $p$-dependent, whereas $A$ is a constant independent of doping [26]. Eqn. 9 incorporates the essential features of a $t$-$J$ model calculation by Prelovsek *et al.*, [20] where inter-plane tunneling has been invoked to describe $c$-axis conductivity of cuprates. It should be noted no explicit term for residual resistivity appears in Eqn. 9. Parameter $\beta_c$ takes into account of the residual resistivity through the tunneling matrix element. Details regarding the modeling of $c$-axis resistivity can be found in Ref. [21].

## 3. Experimental samples and analysis of the resistivity data

High-quality Y123 single crystals were synthesized using the self-flux method in ultra-pure $BaZrO_3$ crucibles. Whereas the 6% Ca doped $Y_{0.94}Ca_{0.06}Ba_2Cu_3O_{7-\delta}$ single crystals were grown using commercial YSZ crucibles with high-purity chemicals. Details of sample preparation, characterization, and measurements of in- and out-of-plane dc resistivities can be found elsewhere [21, 27]. Hole contents were varied by annealing the crystals in different temperatures with different oxygen environments [21, 27]. The $p$-values quoted in this paper, are accurate within ± 0.004. The crystals are detwinned and in-plane resistivity infers to the resistivity along the crystallographic $a$-direction, not affected by the $CuO_{1-\delta}$ chains running along $b$. We show the in-plane dc resistivities of the single crystals in Figs. 1. The out-of-plane resistivities along with the fits to Eqn. 9 are shown in Figs. 2. The in-plane resistivities with respective fits to Eqn. 8 are shown in Figs. 3. A temperature range from $T_c$ + 25 K to 300 K has been used in the fits. The lower temperature limit has been selected such that the resistivity data are not affected by significant SC fluctuations (in other words, significant paraconductive contributions) [28 – 31]. The extracted fitting parameters are shown in Tables 1 and 2. The characteristic PG energy scales (expressed in temperature, $T^*$) and the superconducting transition temperatures are plotted against hole content in Fig. 4. $T_c$ values were located at the zero resistivity point in this study.

## 4. Discussion and conclusions

In the previous section, we have demonstrated that the simple phenomenological model developed in this study accounts for the $p$ and $T$ dependent normal state in- and out-of-plane plane resistivity data remarkably well for double layer pure and Ca substituted Y123 single crystals. The extracted values of the characteristic PG energy scale agree very well with other studies [3, 13, 32]. It is interesting to note that $\varepsilon_g$-values extracted from the analysis of $\rho_p(T)$ and $\rho_c(T)$ data are almost identical. Generally the PG energy scale or equivalently the characteristic PG temperature, $T^*$, is located at the onset of *downward* deviation of the $\rho_p(T)$ data from its high-$T$ linear behavior. While for $\rho_c(T)$, $T^*$ is located at the onset of the *upward* deviation of the data from its high-$T$ linear behavior. These methods consistently yield a higher value of $T^*$ obtained from the $\rho_c(T)$ compared to that obtained from the $\rho_p(T)$ data. The reason being, PG has momentum dependence and $\rho_c(T)$ is dominated by the anti-nodal region of the Brillouin zone where PG is at its most robust. Whereas, the in-plane charge transport is dominated by the nodal part of the Brillouin zone, where PG is weaker [10, 19, 33]. This makes the visual effect of the PG more prominent on the $\rho_c(T)$ data compared to that on the $\rho_p(T)$ data, as seen in Fig. 5. The expression for the thermally averaged EDOS used in this study is also averaged over the $k$-values in the momentum space. Therefore, almost identical values of $\varepsilon_g$ are expected.

Fig. 4 shows that Ca substitution has no noticeable effect on the magnitude of the PG. This agrees completely with previous studies [28, 34 – 36], where the insensitiveness of the characteristic PG energy scale on in- and out-of-plane disorder has been firmly established.

Tables 1 and 2 show that like $\varepsilon_g$, Ca doping has no significant effect on $\alpha_p$ and $\alpha_c$. It is somewhat expected, since $\alpha$ depends mainly on the in-plane hole content. The parameter $\beta_c$, on the other hand, shows both strong $p$ and Ca content dependences. The $p$-dependence can be explained qualitatively mainly by considering the decrease in the $c$-axis tunneling matrix element with underdoping. As far as $c$-axis conduction is concerned, Ca acts as an out-of-plane disorder and is expected to raise the value of $\beta_c$. Parameters $a_p$ and $b_p$ are not independent, both contain the term $\beta_p N_0$. Both $a_p$ and $b_p$ increase with decreasing hole content. Somewhat increased values of $a_p$ and $b_p$ for the Ca doped compounds are probably due to the disordering effect of Ca in place of the Y atoms. Ca resides quite close to the $CuO_2$ planes and may have some effect on in-plane charge dynamics.

At this point we wish to address an issue regarding the residual resistivity of the samples under study. The difference between $a_p$ and $b_p$ gives $\rho_{0p}$, the in-plane residual resistivity. From Table 1, it can be seen that the residual resistivity is negative for all the compounds. The magnitude of this *unphysical* negative residual resistivity increases with underdoping. We believe, it is a consequence of the presence of a PG in the QP energy spectrum. As PG induces a downturn in the in-plane resistivity, any fit incorporating this effect will yield a negative residual resistivity when extrapolated to zero temperature. This is an interesting proposition and worth further attention. Note that for moderately to deeply overdoped cuprates the resistivity becomes superlinear. For such compounds the residual resistivity is always positive [34, 38].

We have confined our analysis for samples with hole content lying within the range from $p = 0.120 - 0.185$ for specific reasons. For deeply UD compounds, the in-plane resistivity develops an upturn at low-$T$ [39], visually similar to the high-$T$ upturn seen in the out-of-plane resistivity. This "localization" effect has not been incorporated within the formalism used in this study. Besides, sample related issues prohibits raising the hole content in Y123 above $p \sim 0.18$; there are reasons to believe the PG vanishes abruptly at $p \sim 0.19$, where the QCP may be located. As mentioned earlier, the proposed model is based on two assumptions – (i) $T$-linearity due to an underlying QCP and (ii) deviation from $T$-linearity due to a PG. Only the range of hole contents where these two assumptions may hold unambiguously have been used for data analysis.

The extrapolated value of extracted $\varepsilon_g(p)$ goes to zero at $p \sim 0.195$ (Fig. 4), quite close to the critical doping as found by a diverse body of experiments on different families of cuprates [13, 25, 28, 34, 37, 38]. Such behavior indicates that PG is not directly related to SC pairing correlations.

Within the proposed scenario, the resistivity anisotropy arises primarily due to the different functional dependence of $\rho_c(T)$ and $\rho_p(T)$ on the thermally averaged EDOS. At high enough temperatures where the EDOS becomes flat (almost constant) the anisotropy ratio, $\rho_c/\rho_p$ becomes temperature independent. Higher the characteristic PG energy scale, higher the onset of the $T$-dependent anisotropy ratio - in complete agreement with experimental observations.

The existence of PG, though poorly understood, is firmly established on experimental grounds. The same cannot be said about the QCP. One of the key point is that there is no clear thermodynamic sign of a QCP related phase transition, as one crosses the $\varepsilon_g(p)$ (or $T^*(p)$) line. Nevertheless, certain versions of orbital current order scenario admit the presence of a QCP without any observable singularity in the specific heat [40]. Within the d-density wave (DDW) scenario, proposals have been made where disorder washes out the thermodynamic signature of the QCP and time reversal symmetry remains the only one that a DDW order can possibly break [41].

In conclusion, we have proposed a minimalistic phenomenological model to describe the doping and temperature evolution of the in- and out-of-plane resistivity for hole doped superconducting cuprates. This simple model provides with a unified scheme to explain the apparently different behavior of $\rho_p(T)$ and $\rho_c(T)$ over a wide range of sample compositions. The proposed model fits the experimental data quite well and the extracted PG energy scale is in very good agreement with earlier findings.

**Acknowledgements**

SHN acknowledges the hospitality provided by the Abdus Salam International Centre for Theoretical Physics (AS-ICTP) Trieste, Italy.

**References**

[1] Leggett A J 2006 *Nature Physics* **2** 134
[2] Norman M R, Pines D and Kallin C 2005 *Advances in Physics* **54** 715
[3] Hashimoto M, Vishik I M, He R-H, Devereaux T P and Shen Z-X 2014 *Nature Physics* **10** 483
[4] Pines D 1997 *Z. Phys. B* **103** 129
[5] Littlewood P and Varma C 1992 *Phys. Rev. B* **46** 405
[6] Anderson P W 1987 *Science* **235** 1196
[7] Alexandrov A S, Kabanov V V and Mott N F 1996 *Phys. Rev. Lett* **53** 2863
[8] Anderson P W 1997 in *The Theory of Superconductivity in High-$T_c$ Cuprates* (Princeton University Press) and the references therein
[9] Dzhumanov D, Ganiev O K and Djumanov Sh S 2014 *Physica B* **440** 17
[10] Su Y H, Luo H G and Xiang T 2006 *Phys. Rev. B* **73** 134510
[11] Kivelson S A, Bindloss I P, Fradkin E, Oganesyan V, Tranquada J M, Kapitulnik A and Howald C 2003 *Rev. Mod. Phys.* **75** 1201
[12] Naqib S H 2012 *Physica C* **476** 10
[13] Tallon J L and Loram J W 2001 *Physica C* **349** 53 and the references therein
[14] Kokanovic I, Cooper J R, Naqib S H, Islam R S and Chakalov R A 2006 *Phys. Rev. B* **73** 184509
[15] Rojo A G and Levin K 1993 *Phys. Rev. B* **48** 16861
[16] Ratan Lal, Ajay, Hota R L and Joshi S K 1998 *Phys. Rev. B* **57** 6126
[17] Misha Turlakov and Leggett A J 2001 *Phys. Rev. B* **63** 064518
[18] Levchenko A, Micklitz T, Norman M R and Paul I 2010 *Phys. Rev. B* **82** 060502
[19] Luo H G, Su Y H and Xiang T 2008 *Phys. Rev. B* **77** 014529
[20] Prelovsek P, Ramsak A and Sega I 1998 *Phys. Rev. Lett.* **81** 3745
[21] Naqib S H, Borhan Uddin M and Cole J R 2011 *Physica C* **471** 1598
[22] Shibauchi T, Carrington A and Matsuda Y 2014 *Anu. Rev. Condens. Matter Phys.* **5** 113
[23] Subir Sachdev 2000 *Science* **288** 475
[24] Hussey N E, Cooper R A, Xiaofeng Xu, Wang Y, Mouzopoulou I, Vignolle B and Proust C 2011 *Phil. Trans. R. Soc.* **369** 1626

[25] Naqib S H and Islam R S 2008 *Supercond. Sci. Technol.* **21** 105017 and references therein

[26] Cooper J R, Minami H, Wittorff V W, Babic D and Loram J W 2000 *Physica C* **341 – 348** 855

[27] Cole J R 2004 *Ph.D. thesis*, University of Cambridge, UK (unpublished)

[28] Naqib S H, Cooper J R, Tallon J L, Islam R S and Chakalov R A 2005 *Phys. Rev. B* **71** 054502

[29] Corson J, Mallozzi R, Orenstein J, Eckstein J N and Bozovic I 1999 *Nature* **398** 221

[30] Naqib S H and Islam R S 2015 *Supercond. Sci. Technol.* **28** 065004

[31] Grbic M S, Pozek M, Paar D, Hinkov V, Raichle M, Haug D, Keimer B, Barisic N and Dulcic A 2011 *Phys. Rev. B* **83** 144508

[32] Lee P A 2008 *Rep. Prog. Phys.* **71** 012501

[33] Xiang T and Hardy W N 2000 *Phys. Rev. B* **63** 024506

[34] Naqib S H, Cooper J R, Tallon J L and Panagopoulos C 2003 *Physica C* **387** 365

[35] Naqib S H, Cooper J R and Loram J W 2009 *Phys. Rev. B* **79** 104519

[36] Naqib S H, Chakalov R A, Cooper J R 2004 *Physica C* **407** 73

[37] Walker D J C, Mackenzie A P and Cooper J R 1995 *Phys. Rev. B* **51** 15653

[38] Cooper R A, Wang Y, Vignolle B, Lipscombe O J, Hayden S M, Tanabe Y, Adachi T, Koike Y, Nohara M, Takagi H, Proust C and Hussey N E 2009 *Science* **323** 603

[39] Ando Y, Komiya S, Segawa K, Ono S and Kurita Y 2004 *Phys. Rev. Lett.* **93** 267001

[40] Grønsleth M S, Nilssen T B, Dahl E K, Stiansen E B, Varma C M and Sudbø A 2009 *Phys. Rev. B* **79** 094506

[41] Sudip Chakravarty, Laughlin R B, Morr Dirk K and Chetan Nayak 2001 *Phys. Rev. B* **63** 094503

**Table 1**

Extracted values of $a_p(p)$ $\alpha_p(p)$, $b_p(p)$, and $\varepsilon_g(p)$ from the fits to $\rho_p(T)$ data.

| Sample | Hole content ($p$) | $a_p(p)$ (mΩ-cm) | $\alpha_p$ (mΩ-cm/K) | $b_p$ (mΩ-cm) | $\varepsilon_g$(K) |
|---|---|---|---|---|---|
| YBa$_2$Cu$_3$O$_{7-\delta}$ | 0.123 | 0.2010 | 0.00101 | 0.307 | 262 |
|  | 0.148 | 0.0251 | 0.00097 | 0.088 | 171 |
|  | 0.164 | 0.0031 | 0.00081 | 0.039 | 122 |
| Y$_{0.94}$Ca$_{0.06}$Ba$_2$Cu$_3$O$_{7-\delta}$ | 0.118 | 0.2730 | 0.00140 | 0.508 | 266 |
|  | 0.122 | 0.2481 | 0.00100 | 0.405 | 254 |
|  | 0.131 | 0.1450 | 0.00096 | 0.306 | 214 |
|  | 0.149 | 0.0692 | 0.00090 | 0.230 | 142 |
|  | 0.169 | 0.0096 | 0.00086 | 0.102 | 107 |
|  | 0.187 | 0.0008 | 0.00060 | 0.046 | 62 |

**Table 2**

Extracted values of $\alpha_c(p)$, $\beta_c(p)$, and $\varepsilon_g(p)$ from the fits to $\rho_c(T)$ data [21].

| Sample | Hole content ($p$) | $\alpha_c$ (mΩ-cm/K) | $\beta_c$ (mΩ-cm) | $\varepsilon_g$(K) |
|---|---|---|---|---|
| YBa$_2$Cu$_3$O$_{7-\delta}$ | 0.123 | 0.0260 | 2.913 | 266 |
|  | 0.148 | 0.0163 | 0.871 | 172 |
|  | 0.164 | 0.0150 | 0.075 | 114 |
| Y$_{0.94}$Ca$_{0.06}$Ba$_2$Cu$_3$O$_{7-\delta}$ | 0.122 | 0.0261 | 4.540 | 282 |
|  | 0.131 | 0.0183 | 4.181 | 230 |
|  | 0.149 | 0.0158 | 3.800 | 171 |
|  | 0.187 | 0.0082 | 2.770 | 12 |

**Figure captions**

Figure 1 (Color online) In-plane resistivity of (a) $Y_{0.94}Ca_{0.06}Ba_2Cu_3O_{7-\delta}$ and (b) $YBa_2Cu_3O_{7-\delta}$. Hole contents are shown in the plots.

Figure 2 (Color online) Out-of-plane resistivity and respective fits to Eqn. 9 (full black lines) for (a) $Y_{0.94}Ca_{0.06}Ba_2Cu_3O_{7-\delta}$ and (b) $YBa_2Cu_3O_{7-\delta}$. Hole contents are shown in the plots. For clarity one in ten experimental data points are shown only (reproduced from Ref. 21).

Figure 3 (Color online) In-plane resistivity and respective fits to Eqn. 8 (full black lines) for (a) $Y_{0.94}Ca_{0.06}Ba_2Cu_3O_{7-\delta}$ and (b) $YBa_2Cu_3O_{7-\delta}$. Hole contents are shown in the plots. For clarity one in ten experimental data points are shown only.

Figure 4 (Color online) Characteristic pseudogap energy (in K) and superconducting transition temperature of $Y_{0.94}Ca_{0.06}Ba_2Cu_3O_{7-\delta}$ and $YBa_2Cu_3O_{7-\delta}$. The dashed-dotted line shows the $T_c(p)$ trend for $Y_{0.94}Ca_{0.06}Ba_2Cu_3O_{7-\delta}$.

Figure 5 (Color online) Conventional method of extraction of the pseudogap temperature from in- and out-of-plane resistivity data. The arrows locate $T^*$ of $YBa_2Cu_3O_{7-\delta}$ ($p = 0.148$).

Figure 1

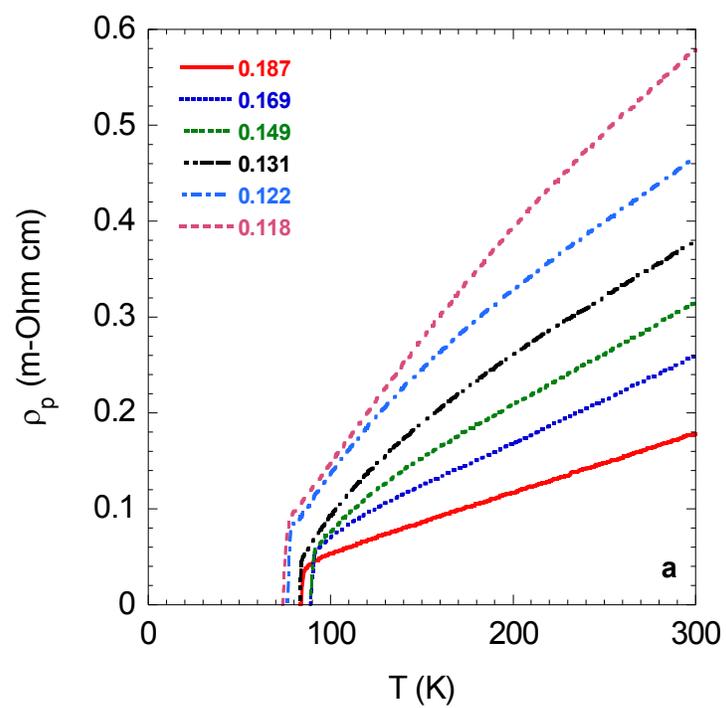

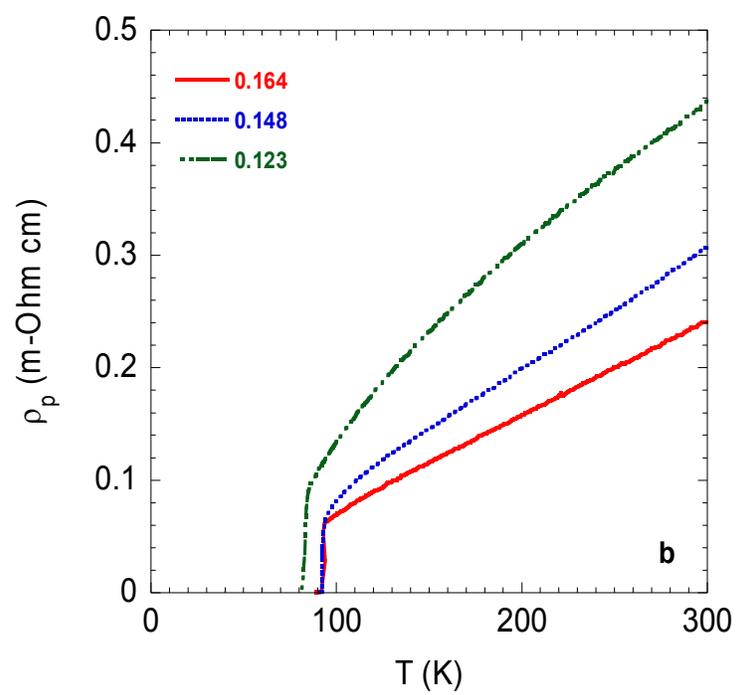

Figure 2

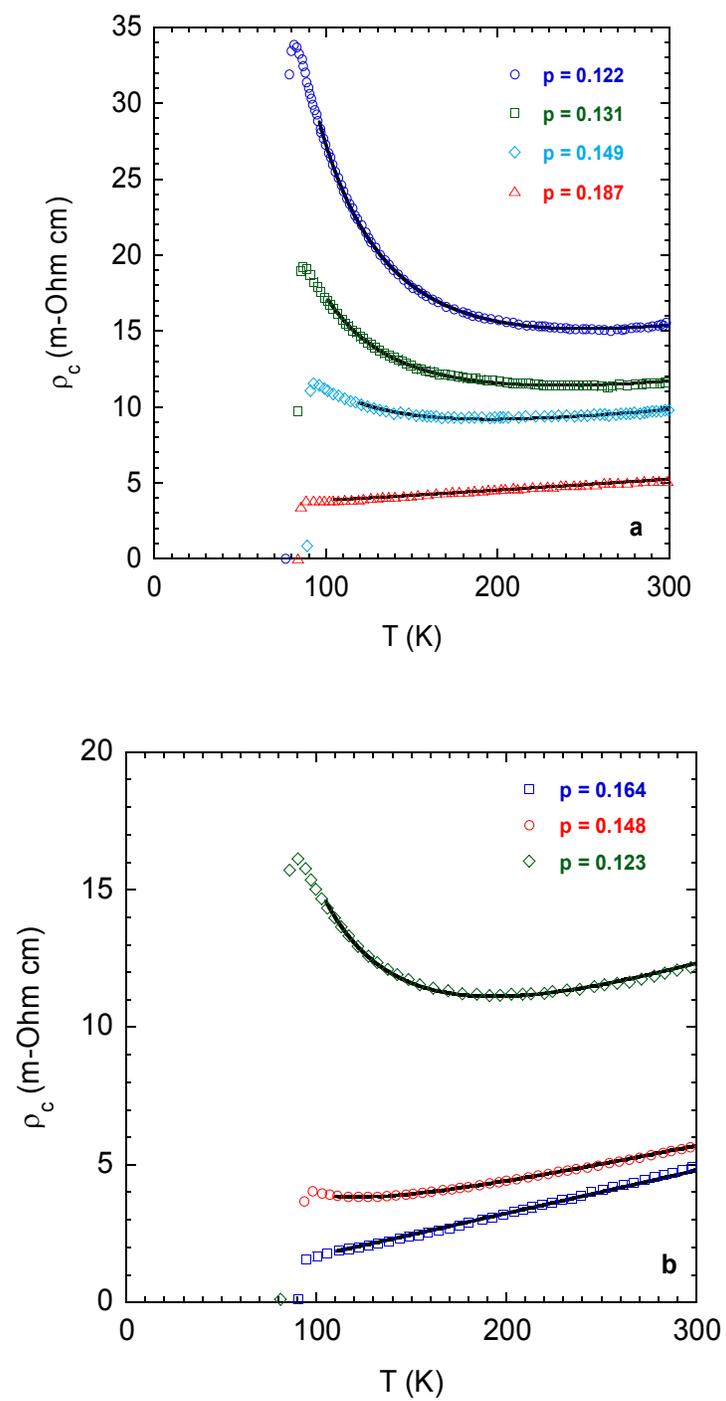

Figure 3

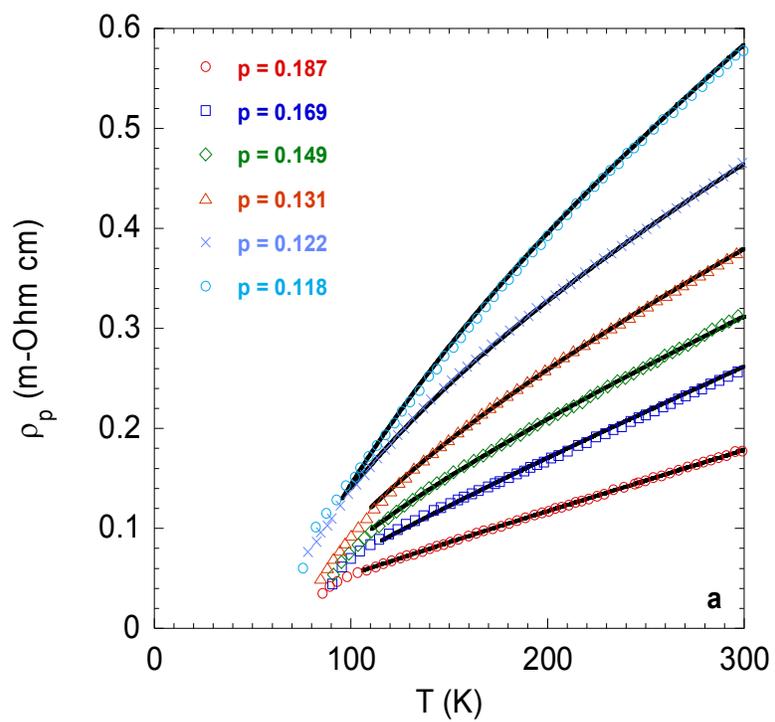

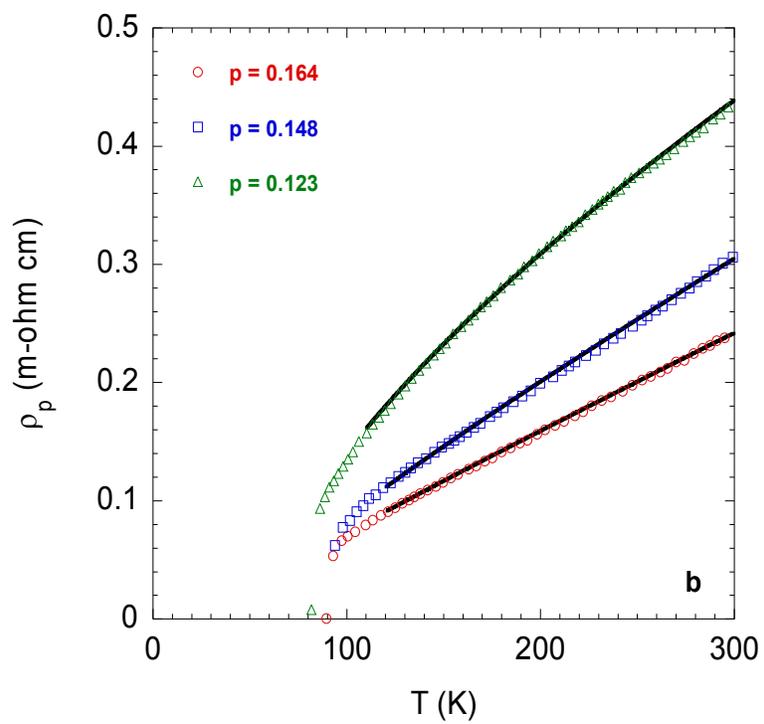

Figure 4

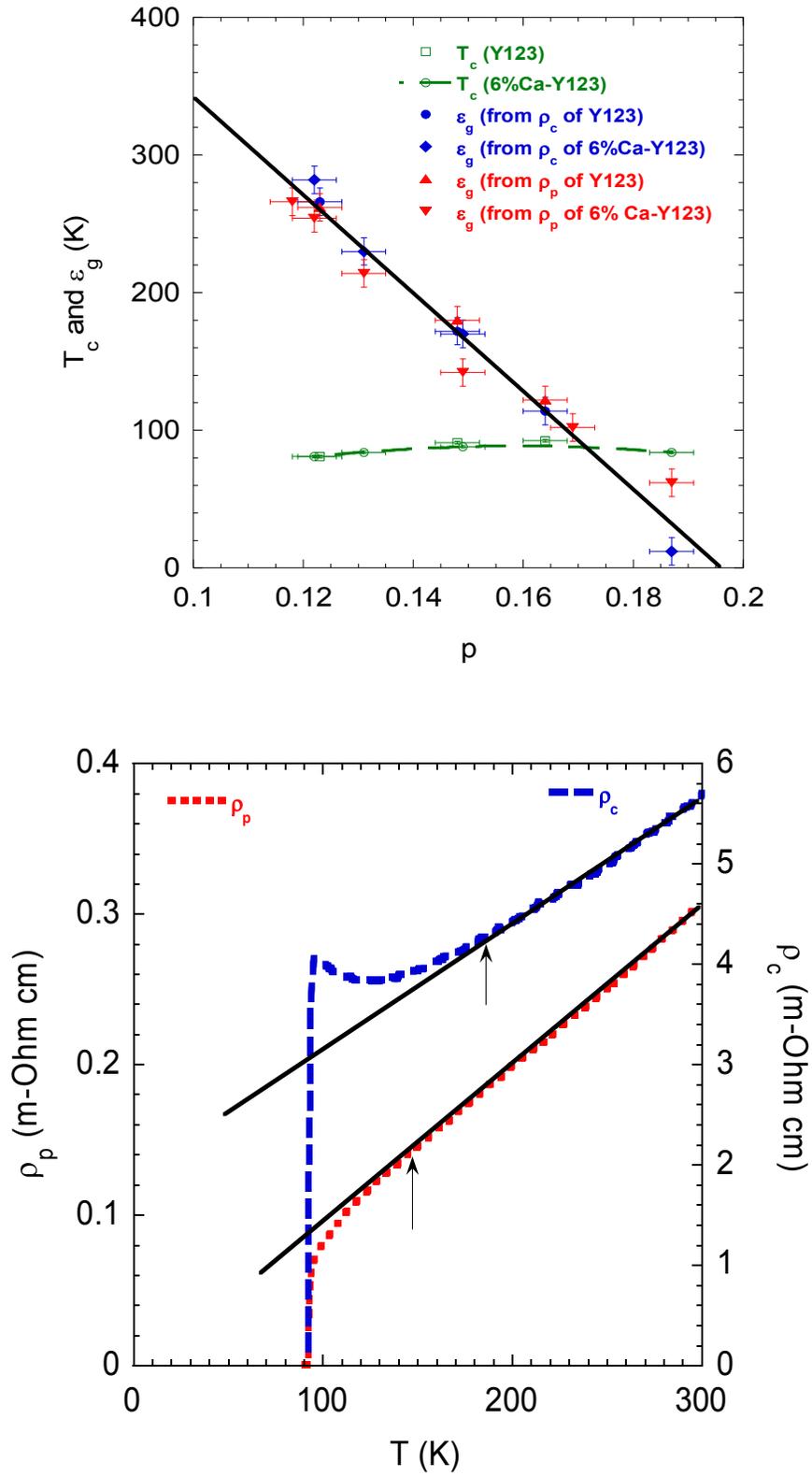